\documentclass[12pt]{iopart}

\usepackage{graphicx}
  
\begin{document}

\title{Propagation of ultra high energy cosmic rays}

\author{Todor Stanev}

\address{Bartol Research Institute and Department of Physics and Astronomy,
University of Delaware, Newark DE19716, U.S.A.}
\ead{stanev@bartol.udel.edu}
\begin{abstract}
 We briefly describe the energy loss processes of ultrahigh energy
 protons, heavier nuclei and gamma rays in interactions with
 the universal photon fields of the Universe. We then discuss
 the modification of the accelerated cosmic ray energy spectrum
 in propagation by the energy loss processes and the charged
 cosmic ray scattering in the extragalactic magnetic fields.
 The energy lost by the ultrahigh energy cosmic rays goes
 into gamma rays and neutrinos that carry additional 
 information about the sources of highest energy particles.
 The new experimental results of the HiRes and the Auger 
 collaborations are discussed in view of the predictions 
 from propagation calculations.  
\end{abstract}
\maketitle

\section{Introduction}

 The high interest of the astrophysical community to the
 Ultrahigh Energy Cosmic Rays (UHECR) developed after the Agasa 
 experiment showed its energy spectrum that continued above
 10$^{20}$ eV without a break~\cite{Agasa} and with no indication
 of a GZK~\cite{GZK} structure. This is not the first time cosmic
 rays of such high energy were observed. John Linsley reported
 on the first extensive air shower above 10$^{20}$ eV in 1963~\cite{Linsley}.
 Three years later Greisen, and independently Zatsepin \& Kuzmin,
 predicted the {\em end of the cosmic ray spectrum} that results
 from the interactions of the UHECR with the microwave background
 radiation (MBR). Every giant air shower experiment has reported
 events exceeding 10$^{20}$ eV since that time. Because of the 
 very low flux of such particles and the various energy estimates
 of different experiment it was impossible to judge if these
 highest energy nuclei in the Universe obey the GZK predictions - 
 a sharp decline of the spectrum above 4$\times$10$^{19}$ eV or not.
 At the time Agasa was the highest exposure experiment and it claimed 
 a spectrum that is not consistent with the GZK {\em cut-off}.

 This result was not confirmed by the contemporary leaders in
 UHECR statistics, the Auger and the HiRes experiments.
 Both these groups have published UHECR spectra that seems to be
 consistent with the GZK prediction. There are, however, distinct
 differences between the results of these two groups that 
 create new challenges for the scientists who are eager to 
 establish in some detail the acceleration and propagation 
 picture of UHECR. 

 It has been a common suspicion for more than 50 years that
 the highest energy cosmic rays are of extragalactic origin~\cite{Cocconi}.
 The argument is that our Galaxy is not large enough and the
 galactic magnetic fields are not strong enough to contain 
 particles of such high energy and thus accelerate them. 
 The gyro radius $R_g$ of a 10$^{20}$ eV proton in 3$\mu$G field 
 is higher than 30 kpc, similar to the dimension of the whole 
 Galaxy that does not contain shocks of the same dimension.

 If UHECR are indeed of extragalactic origin they should suffer 
 from the interactions with the MBR unless their sources are 
 cosmologically very close to us. For this reason a comparison 
 of the detected UHECR spectrum with theoretical calculations
 of the particle propagation in extragalactic space should at
 least point at the distance distribution of their sources and
 bring us closer to the source identification. From that point
 of view the small differences between the HiRes~\cite{HiRes_prl} 
 and Auger~\cite{Auger_prl} spectra lead us to very different
 interpretations.

\begin{figure}[thb]
\centerline{\includegraphics[width=8truecm]{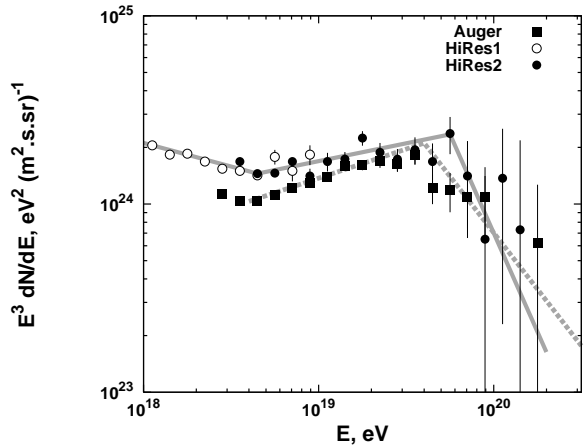}}
\caption{ The energy spectra of Auger and HiRes recently published 
 in Phys. Rev. Letters. The thick gray lines show the spectral
 fits presented in the same papers.}
\label{Aug_HR_prl}
\end{figure}

 Figure~\ref{Aug_HR_prl} shows the published spectra and the rough 
 fits of the observed spectra that are suggested by the two
 experimental groups. It is obvious that the extremely small 
 statistics around 10$^{20}$ eV makes the fit of the strong 
 decline above 10$^{19.6}$ eV very difficult. The statistics 
 around 10$^{18.5}$ eV, on the other hand, is significant and
 the different behavior of the spectra is confusing in addition
 to the different normalizations of the spectra. 
 Another major difference between the two experiments is their
 estimate of the chemical composition of UHECR studied from the
 measured depth of shower maximum $X_{max}$. HiRes derives
 a pure proton composition~\cite{HR_comp}, while 
 the Auger analysis tends to reveal a more complex mixed
 nuclear composition~\cite{Auger_Unger}. The UHECR composition
 at their sources affects the interpretation of the spectra 
 because protons and heavier nuclei have different energy
 loss mechanisms.
  
 This paper is organized as follows: Section 2 discusses 
 the energy loss of different possible primary particles. 
 The next section describes the formation of the ultrahigh
 energy cosmic ray spectrum on propagation and the astrophysical
 parameters that are important for it. Section 4 is dedicated
 to the production of secondary particle fluxes in propagation,
 and mostly to cosmogenic neutrinos. Section 5 contains a brief
 summary of the current knowledge.

\section{UHECR energy loss}

 Apart from the adiabatic energy loss due to the expansion of
 the universe there are two important processes for protons:
 photoproduction interactions and $e^+e^-$ pair production (BH)
 interactions identical to the pair production interactions
 of $\gamma$--rays in the nuclear field. 
 The back of the envelope estimate of the photoproduction energy
 loss goes like this:
 The average interaction length $\lambda_{ph}$
 for interactions with the MBR
 is the inverse of the product of the interaction cross section
 $\sigma_{ph}$ and the photon density $n$. For $\sigma_{ph}$ =
 10$^{-28}$~cm$^2$ and $n$ = 400~cm$^{-3}$ $\lambda_{ph}$ = 8.3~Mpc.
 Since protons lose about 0.2 of their energy in each interaction
 it takes about ten interaction length to decrease the particle
 energy by a factor of 10. 

 The story of the heavier nuclei energy loss is more complicated.
 In addition to these two processes heavy nuclei lose energy in
 photo disintegration (spallation) processes, i.e. when the
 center of mass energy exceeds the giant dipole resonance the
 nucleus can lose a nucleon. Since less energy is required 
 in the center of mass the cross section is higher but
 the energy loss depends on the mass of the nucleus that loses
 only one or two nucleons. The photoproduction energy loss follows the
 same energy dependence as for protons but in the Lorentz factor
 space, i.e. in E/A units. The pair production cross section 
 is a quadratic function of the charge of the nucleus $Z$. 

 In the case of exotic theoretical {\em top-down} models UHECR
 are gamma rays from the decay of very heavy $X$--particles.
 Although the $\gamma$--ray fraction was strongly limited by
 the Auger Collaboration to not more than 2\% for UHECR above
 10$^{19}$ eV \cite{Auger_gamma} this possibility can not be
 totally excluded. In such a case the energy loss is due to the
 $\gamma \gamma \rightarrow e^+e^-$ process.
 
\subsection{Proton energy loss length}

 A photoproduction interaction is possible when at least one pion is
 generated in the process. This requires that the center of mass
 energy of the interaction $\sqrt{s}$ is higher that the sum of a
 proton mass $m_p$ and a pion mass $m_\pi$. In the laboratory system
 the square of the center of mass energy $s$ is
\begin{equation}
 s \; = \; m_p^2 + 2 E_p \epsilon (1 - \cos{\theta})\; ,
\end{equation}
 where $\epsilon$ is the photon energy and $\theta$ is the angle between
 the proton and the photon. In a head on collision ($\cos{\theta} = -1$)
 with a photon of the average MBR energy (6.3$\times$10$^{-4}$~eV) the
 minimum proton energy is
\begin{equation}
E_p \; = \; \frac{m_\pi}{4 \epsilon}(2 m_p + m_\pi) \; \simeq \;
 {\rm 10^{\rm 20}\; eV.}
\end{equation}
 There are many MBR photons with higher energy and the threshold 
 proton energy is actually lower, about 3$\times$10$^{19}$ eV.

 The cross section for this interaction is very well studied at accelerators
 where photons interact with stationary protons. The highest cross section 
 is at the mass of the $\Delta^+$ resonance (1232 MeV) which decays to
 either a proton and a neutral pion ($p \pi^0$) or to a neutron and a 
 positive pion ($n \pi^+$). At the peak of the resonance the cross section 
 is about 500 $\mu$b. At higher energy the cross section first decreases
 to about 100 $\mu$b and then increases logarithmically. The neutron 
 interaction cross section is, if not identical, very similar to the
 proton one.

 The MBR spectrum and density are also very well known, so the proton 
 interaction length can be calculated exactly, as shown in the left hand
 panel of Fig.~\ref{xlosses} with dash line. Since protons lose only
 a fraction of their energy $(K_{inel})$, another quantity - the energy loss
 length $L_{loss} \; = \; - \frac{1}{E} \frac{dE}{dx}$ becomes important.
 The energy loss length is longer than the interaction length by 
 $1/K_{inel}$, by about a factor of 5 at threshold. At higher energy 
 $K_{inel}$ grows and this factor is about 2. 

 In the case of $e^+e^-$ pair production~\cite{BerGri}
 the addition of two electron masses to the center of mass energy $\sqrt{s}$
 requires much lower proton energy and the process has lower threshold.
 The cross section for pair production is higher than $\sigma_{ph}$,
 but the fractional energy loss is small, of order of the ratio of the
 electron to proton mass $m_e/m_p$. The energy loss length has a minimum
 around 2$\times$10$^{19}$~eV and is always longer than 1,000 Mpc.

 The last proton energy loss process is the redshift due to the expansion
 of the Universe. The current energy loss length to redshift is the
 ratio of the velocity of light to the Hubble constant ($c/H_0$) and is
 4,000 Mpc for $H_0$ = 75 km.s$^{-1}$Mpc$^{-1}$.  
     
  The energy loss of protons in the MBR is shown in the left hand
 panel of Fig.~\ref{xlosses}. This figure shows also the 
 photoproduction interaction length and the decay length of
 neutrons. The neutron photoproduction cross section is almost
 identical to the proton one. This means that neutrons of energy
 less than 4$\times$10$^{20}$  will most likely decay and only
 neutrons of higher energy are likely to have
 photoproduction interactions.

 \subsection{Energy loss length of heavier nuclei}

\begin{figure}[thb]
\includegraphics[width=7.5truecm]{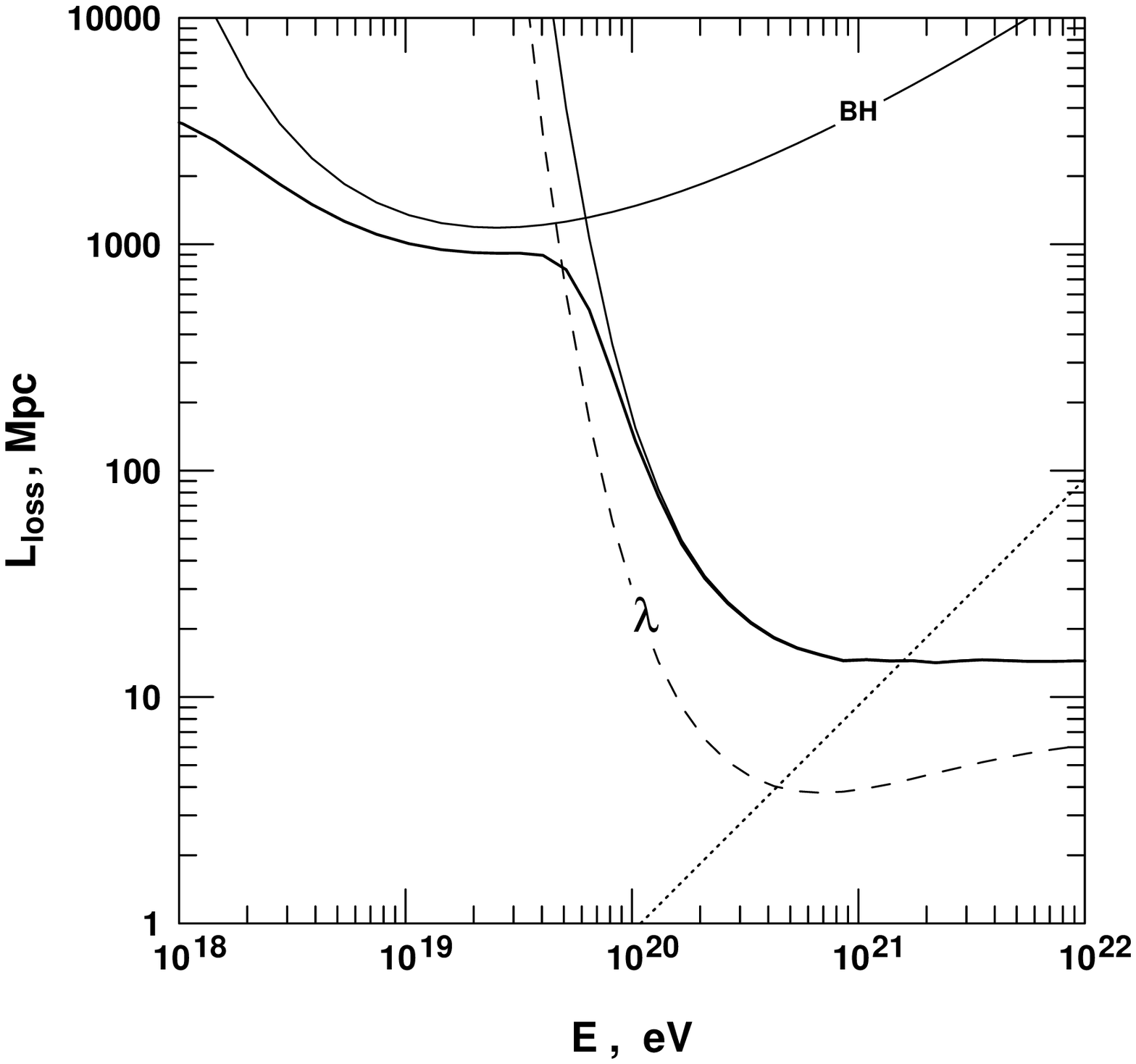}
\includegraphics[width=7.5truecm]{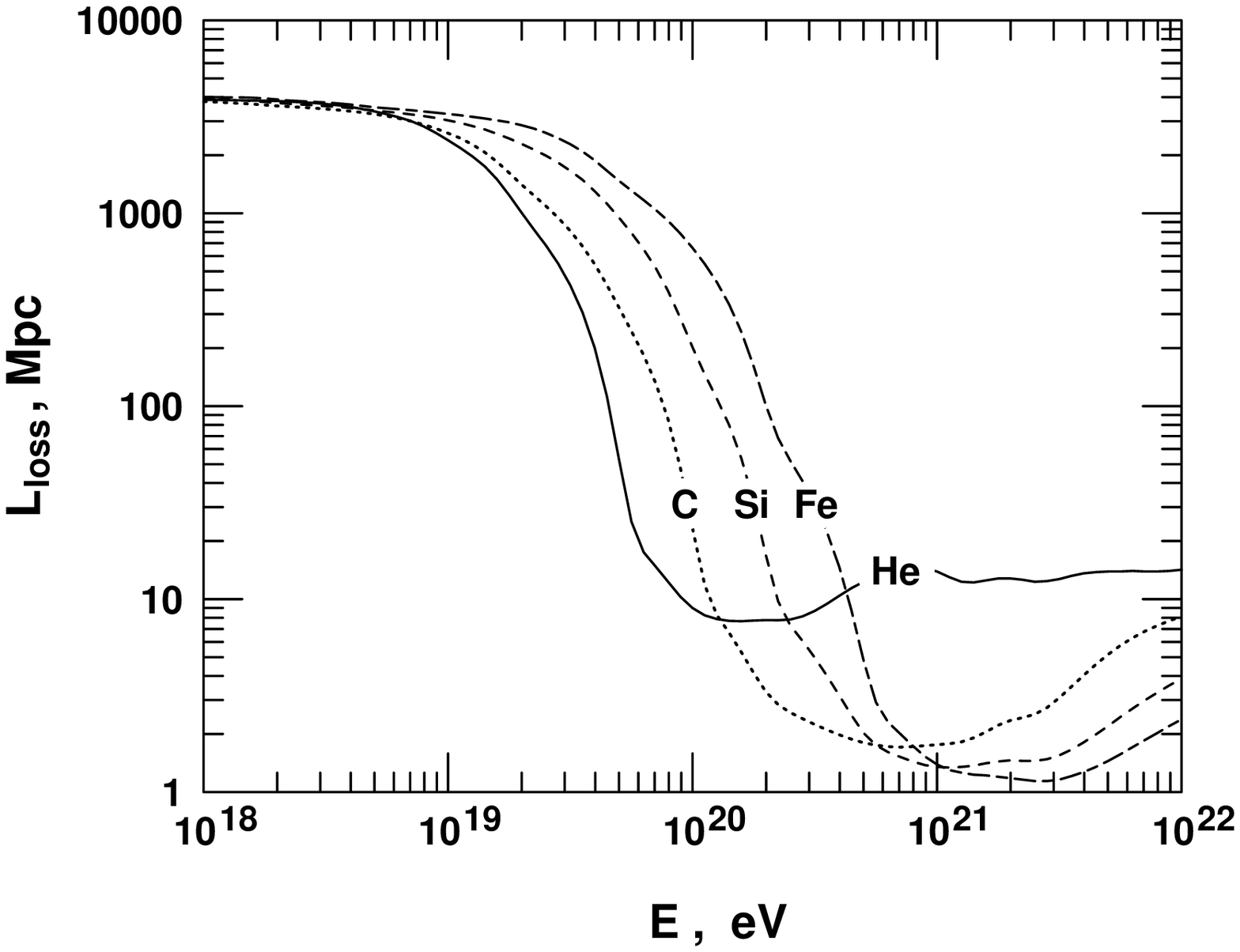}
\caption{Left hand panel: interaction length and energy loss length
 for protons in the MBR and neutron decay length. Right hand panel:
 energy loss length for different nuclei as calculated in 
 Ref.~\protect\cite{bertoneetal}.}
\label{xlosses}
\end{figure}

 The energy loss length of heavier nuclei is shown in the
 right hand panel of Fig.~\ref{xlosses} as calculated in
 Ref.~\cite{Bertoneetal}. Its minimum value is significantly 
 lower than that of protons but is achieved at higher energy:
 $A\times E_p$. It is, of course, not obvious that
 such high energy could be achieved at UHECR acceleration,
 although we always assume that the maximum acceleration energy
 is proportional to the charge $Z$ of the nucleus.
 The effect of propagation on the accelerated UHECR can not
 be calculated directly from the energy loss lengths shown in
 Fig.~\ref{xlosses} because an accelerated nucleus changes its mass
 after the first photo disintegration interactions. This means
 that the code that treats propagation of nuclei should be able
 to calculate the cross sections for all nuclei and isotopes
 lighter than the injected nucleus. In addition to the cross
 sections for losing one, two, and more nucleons the codes should 
 evaluate the decay probability for unstable nuclei that may
 be generated in the propagation.

 Many of the cross sections necessary for a correct simulation
 of the propagation of nuclei have been experimentally studied
 at accelerators. There is still a need for approximations for
 some of the numerous processes taking place in the 
 propagation of nuclei. See, e.g. the early work on propagation
 of nuclei~\cite{PSB76} that is based on such approximations.
 
\subsection{Gamma ray energy loss length}

  The process has $\gamma \gamma \rightarrow e^+e^-$ has a
 resonant character and the cross section peaks at $E_\gamma \epsilon \; = \;
 2 m_e^2$, where $\epsilon$ is the energy of the seed photons. 
 For the average energy of MBR this corresponds to $E_\gamma$ of
 8$\times$10$^{14}$~eV and the mean free path decreases with increasing
 $E_\gamma$. For gamma rays of energy 10$^{20}$~eV the relevant seed
 photon frequency is about 1 MHz - in the radio band. 
 This creates a big uncertainty in the estimates of the UHE $\gamma$-ray
 energy loss length because the density of the radio background at such
 frequencies is not known. One can relate the radio emission of various
 astrophysical objects to the much better known infrared emission and
 generate models to calculate the energy loss length. The results of
 such modeling are energy loss lengths of order that of protons.
 
 A different source of uncertainty in the $\gamma$-ray propagation
 is the strength of the extragalactic magnetic fields. If they are
 negligible the electrons have inverse Compton interactions,
 whose interaction length is similar to that of the pair production,
 and generate a second generation of very high energy $\gamma$-rays.
 This cascading can continue for a significant
 distance without downgrading very much the gamma ray energy. If, however,
 the magnetic fields are significant electrons lose energy very fast on
 synchrotron radiation. The $\gamma$-ray energy is rapidly transferred
 to the MeV-GeV energy range. The range of top-down cosmic ray generation
 models has been restricted because of overproduction of GeV $\gamma$-rays. 
 The energy loss distance on synchrotron
 radiation is 2.6$E_{18}^{-1} B_{-9}^{-2}$~Mpc, where $E_{18}$ is the
 electron energy in unites of 10$^{18}$~eV and $B_{-9}$ is the strength
 of magnetic field in nGauss. 

\section{Formation of the cosmic ray energy spectrum after propagation}

 Predictions of the shape of the cosmic ray spectrum requires much
 more than the energy loss in propagation. The necessary astrophysical
 input, currently unknown,
 includes et least the following four items:\\
 \ \ \ \ $\bullet$ UHECR source distribution\\
 \ \ \ \ $\bullet$ cosmic ray source luminosity\\
 \ \ \ \ $\bullet$ cosmic ray injection (acceleration) spectrum\\
 \ \ \ \ $\bullet$ maximum acceleration energy E$_{max}$\\
 \ \ \ \ $\bullet$ cosmic ray chemical composition\\
 \ \ \ \ $\bullet$ cosmic ray source cosmological evolution.

 The UHECR source distribution was the least known one.
 After Auger found a correlation with nearby (redshift $z$ less than
 0.018) active galactic nuclei (AGN)of their highest energy
 events~\cite{Auger_sci,Auger_corr}
 the situation changed. Although there is no certainty that these
 AGN are indeed the sources of UHECR, this is the first time we
 have a suggestion 
 what the source distribution could be.
  
 The other five parameters are not independent of each other. The UHECR
 source luminosity can in principle be determined by the detected UHECR flux
 above 10$^{19}$ eV. In view of the low current statistics, the derived
 luminosity depends strongly on the assumed injection spectrum and 
 composition and partially
 on the assumed cosmological evolution of the sources~\cite{ddmts}. The source
 cosmological evolution may be the best known parameter since it should
 resemble these of other astrophysical phenomena such as the star formation
 rate in the Universe.

 \subsection{Formation of the proton spectrum in propagation}

\begin{figure}[thb]
\includegraphics[width=7.5truecm]{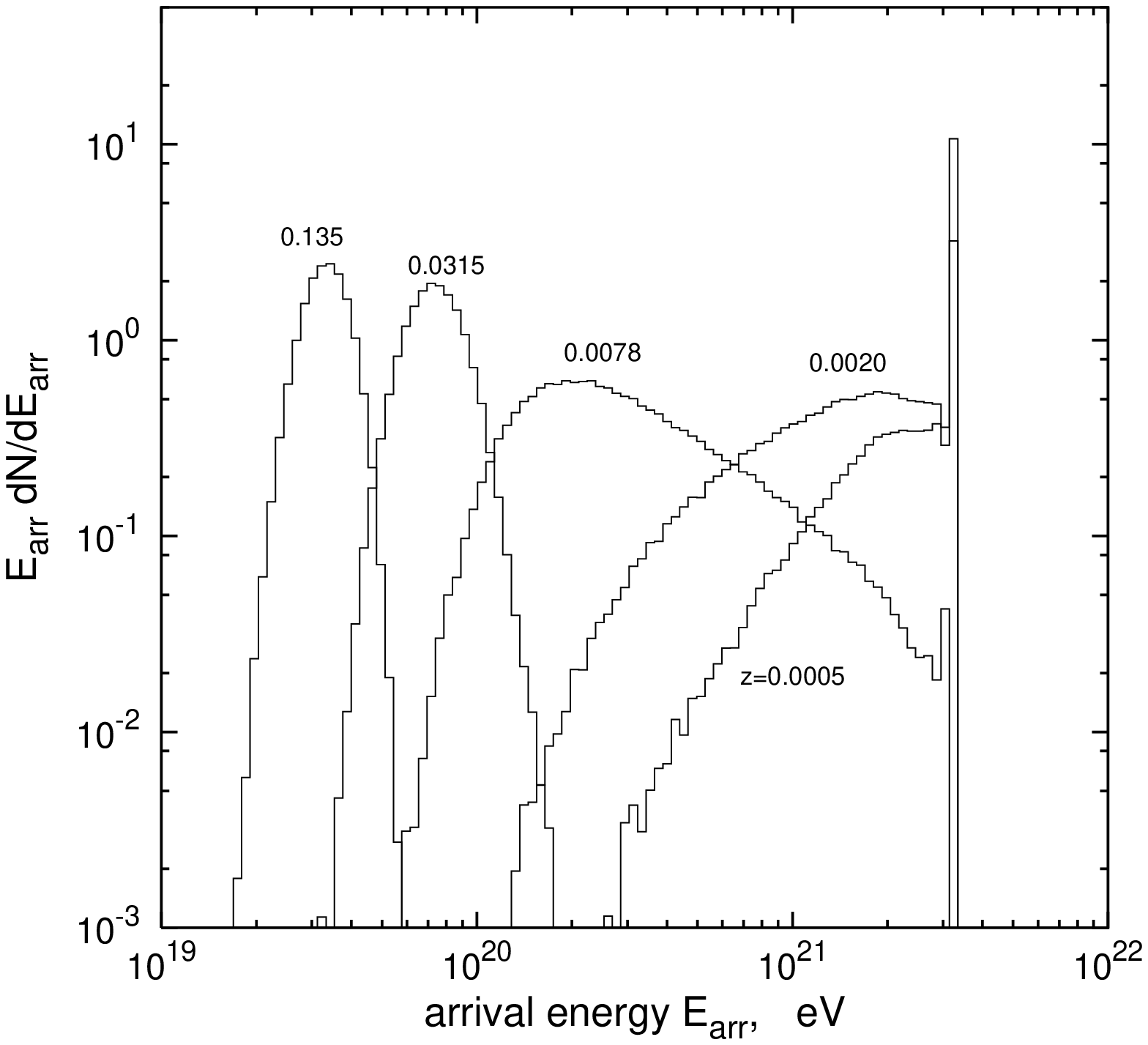}
\includegraphics[width=7.5truecm]{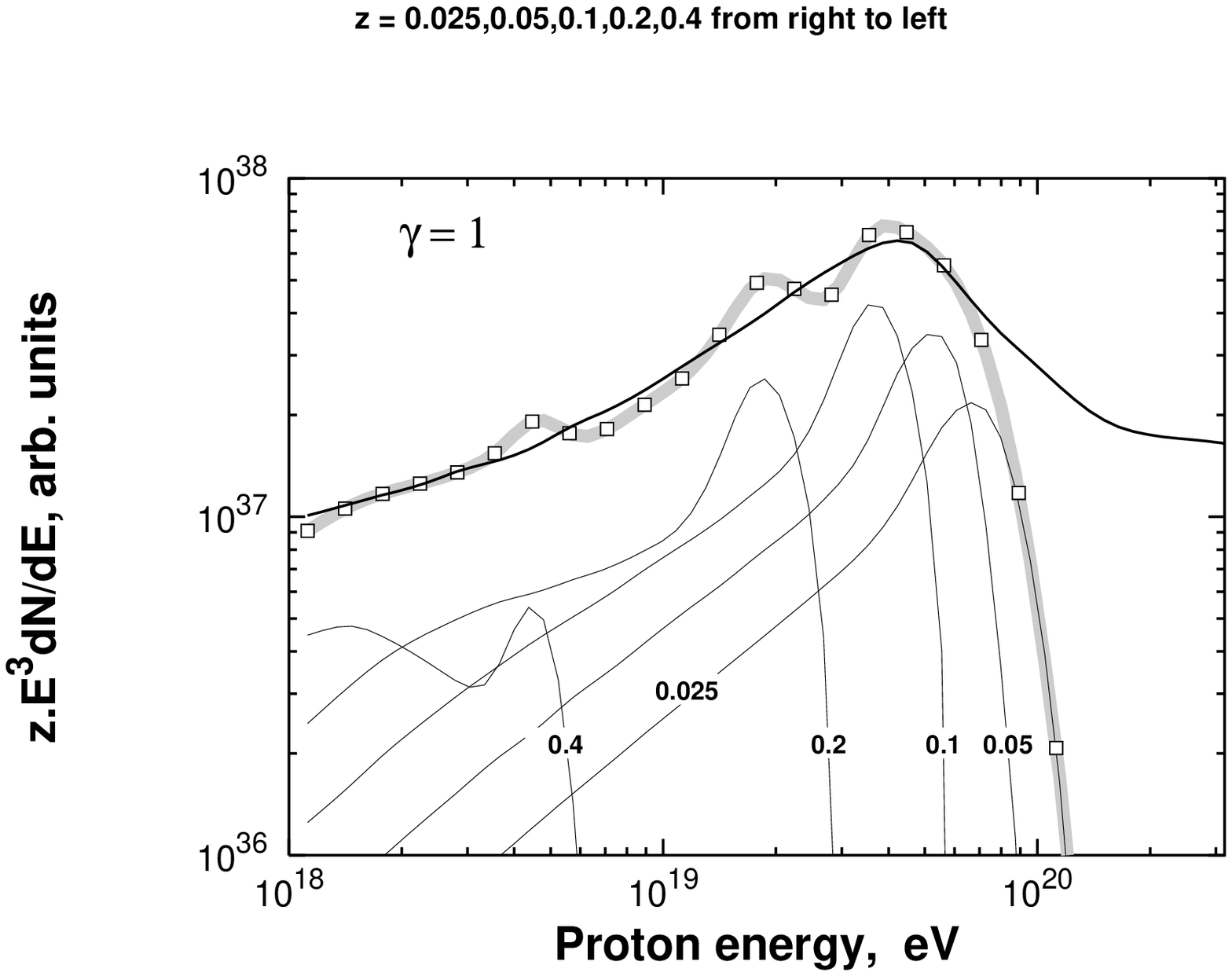}
\caption{Left hand panel: Arrival energy distribution of 10$^{21.5}$ eV
 protons after propagation from different redshifts that are indicated
 by the distributions. Right hand panel: contribution of different 
 redshifts to the arrival spectrum for $E^{-2}$ injection spectrum
 with no cosmological evolution. The thick gray line shows the sum
 of the contributions from these five redshifts while the black
 line is result of a full integration.
 }
\label{prop} 
\end{figure}

 As an example we will discuss the formation of the cosmic ray spectrum
 in the case all UHECR are protons. The left hand panel of Fig.~\ref{prop}
 shows the evolution of a monochromatic protons of injection
 energy 10$^{21.5}$ eV after propagation from different redshifts.
 After propagation on $z$=0.0005 (appr. 2 Mpc) a large fraction
 (appr. 60\%) of the injected protons have not interacted. 
 Some of them, however, have interacted a couple of times and
 their energy has decreased as much as a factor of 10 as such high
 energy protons lose much more than 20\% of their energy in
 photoproduction interactions. 
 Almost all protons have interacted at $z$=0.0078 and the average
 proton energy at this distance is 2$\times$10$^{20}$ eV.
 In further propagation the energy distribution at arrival 
 becomes narrower as the highly stochastic photoproduction 
 energy loss becomes lower than the almost continuous pair production,
 and later the adiabatic loss from the expansion of the Universe.
 The width of the arrival energy decreases until the pair production
 energy loss length is smaller than that of the adiabatic energy 
 loss length.   
 One can see in the right hand panel how the sums of the contributions
 of low redshifts exceeds the value of the injected spectrum after 
 a propagation to $z$=0.1. In models with the cosmological evolution
 the effect is stronger and proportional to the strength of the
 source evolution.

 A natural assumption for the source distribution, the result of which
 is shown in Fig.~\ref{prop} is that sources are 
 isotropically and homogeneously distributed in the Universe because
 we do not inhabit a special part of it, and the contribution of
 all sources are identical. In such a case the cosmic ray
 flux at Earth could be determined by an integration of the fluxes
 from different redshifts
 shown in the right hand panel of Fig.~\ref{prop}. In the case of 
 cosmological evolution of the sources the integral is
\begin{equation}
 N(E) \; = \; \int^{z_{max}}_0 \int_E^{E_0} L(z) N_0(E_0) P(E_0, E', z) 
 \frac{dt}{dz} dE' dz \; ,
\end{equation}
 where $L(z)$ is the cosmic ray source luminosity as a function of redshift
 and $N_0(E_0)$ reflects the injection spectrum, $P(E_0,E',z)$ is the
 probability for a proton injected with energy $E_0$ at redshift $z$ to
 reach us with energy $E'$.
 The derivative $dt/dz$ depends on the cosmological model
 and is
 $$ {{dt} \over{dz}} \; = \; {{1} \over {H_o (1 + z)}}
 [ \Omega_M (1+z)^3 + \Omega_\Lambda ] ^{-1/2}$$
 and is simplified to $(1+z)^{-5/2}/(H_o(1 + z))$ for the
 Einstein-deSitter Universe.
 
 It is important to note that contribution of different redshifts
 depends not only on the cosmological evolution but also on the 
 injection spectral index as the photoproduction energy loss is
 a strong function of the injection energy. Since in steep 
 injection spectra a larger fraction of the observed flux 
 comes from lower primary energy (that do not change as much
 on propagation) the contribution of higher redshifts is larger.
 
 One can see in the right hand panel of Fig.~\ref{prop} that
 even $z$=0.05 does contribute to UHECR above 6$\times$10$^{19}$ eV
 which was observed by Auger to be correlated to AGN within 
 distances of only 71 Mpc ($z \leq$ 0.017). There are several ways
 to deal with 
 this seeming controversy. One would be to claim that the energy
 scale of Auger is not correct and it should be higher by about 25-30\%. 
 This would make the contributions of higher redshifts smaller
 and bring the GZK sphere closer to the expectations for protons.

 Another is to claim that the feature observed by Auger and HiRes is not
 a result of the GZK process, as the experimental groups claim, and is just
 the end of the acceleration
 power of the sources that does not much exceed 10$^{20}$ eV. The
 shape of the spectra observed by both experiments (that are
 shown in Fig.~\ref{Aug_HR_prl}) is, however, 
 very close to what we expect from the GZK cutoff.
 If one attempts to fit these spectra with the same model some
 problems start showing up. The model of Berezinsky et al~\cite{Beretal}
 that has pure proton composition with steep injection
 spectrum ($E^{-2.7}$) fits perfectly the HiRes spectrum
 and does not require other contributions (such as galactic cosmic
 rays) above 10$^{18}$ eV. 
 The same model, however, does not appear to fit well the
 current Auger data.

 One can also involve the extragalactic magnetic fields in the
 explanation. If it is high the UHECR would scatter
 often and their real pathlength would be considerably larger 
 than the distance to the sources.

\subsection{Propagation in magnetic fields} 

 Our knowledge of the extragalactic magnetic fields is quite 
 limited. We do know that they exist and are most likely 
 proportional to the mass density in the Universe, i.e.
 high in regions of high matter concentration ans low in voids.
 Magnetic fields introduce three main effects. The cosmic ray
 scattering increases the propagation pathlength and thus
 restricts the radius of the possible sources. The scattering
 creates a deviation of the arrival direction of the UHECR
 from the direction of the source. The gyro radius of a 10$^{20}$~eV
 proton in 10$^{-9}$ G (nG) field is 100 Mpc. 
 If the field is random with a correlation length $\ell$
 the deviation angle $\langle \theta \rangle$ after propagation on
 distance $D$ is
\begin{equation}
 \langle \theta \rangle \; \simeq \; {\rm 2.5}^\circ
 B_{-9} D_{100}^{\frac{1}{2}} \ell_1^{\frac{1}{2}} E_{20}^{-1} \; ,
\end{equation}
 where $B_{-9}$ is the r.m.s. field strength in nG, $D_{100}$
 is the distance in units of 100 Mpc, $\ell_1$ is the correlation
 length in units of 1 Mpc and $E_{20}$ is the energy in units of
 10$^{20}$~eV. Protons below 10$^{20}$ eV would scatter much more
 around the direction of the source but the highest energy particles
 would point at the source with an angle comparable to the experimental
 resolution. The scattering angles for heavier nuclei are 
 proportional to their charge. 

 The scattering also introduces time delay compared to the rectilinear
 propagation of light. The time delay $\delta \tau$ has a much
 stronger dependence
 on the particle energy, magnetic field strength and
 the propagation distance. For small angle scattering it is
\begin{equation}
 \delta \tau \; \simeq \; {\rm 3} \times {\rm 10}^5
 B_{-9}^2 D_{100}^2 \ell_1 E_{20}^{-2}\; {\rm years}.
\label{delay}
\end{equation}
 If the source of the observed UHECR were an explosive process,
 such as a gamma ray burst at a distance of 100 Mpc, all protons
 would accelerated at once, but because of time delay they would
 arrive at Earth in a reverse order of their energy. The highest
 energy particles would reach us first, while the lower energy ones 
 would be delayed with millions of years. It is important to note that
 the time delay depends on the square of the particle charge. 
 Iron nuclei coming from 10 Mpc will be still delayed
 seven times more than protons arriving from 100 Mpc.

 Time delays could prevent some of the extragalactic protons 
 from reaching us, because their travel time could exceed
 the the age of the Universe. Particles of energy below
 5.5$\times$10$^{17}$ eV from the gamma ray burst at 100 Mpc,
 for example, propagating in 1 nG field will not reach us
 because their time delay will
 exceed Hubble time, taken here of 10$^{10}$ yrs for simplicity.

 In a simulation that propagated protons in 1 nG random
 field~\cite{Staetal00} we calculated the {\em proton horizon}
 $R_{50}$, which is the distance at which $1/e$ fraction of
 protons maintain at least one half of their injection
 energy. This distance is smaller than the proton energy loss
 length even for 10$^{20}$ eV protons. The ratio  $R_{50}$ to
 the energy loss length in 1 nG field is proportional to $E^{-1.2}$
 at the approach to 10$^{20}$ eV. Using Eq.~\ref{delay} one
 can estimate the additional pathlength in 1 nG field for
 6$\times$10$^{19}$ eV (60 EeV) protons emitted at distance of
 70 Mpc to 66 kpc. This does not make a huge difference but
 remember that in 10 nG field it would grow to 6.6 Mpc. For
 iron and 1 nG field the additional pathlength would increase to
 45 Mpc - for a total pathlength of 115 Mpc.

 The numbers quoted above do not include the cosmic ray energy loss
 which contributes significantly to the low $R_{50}$ values.
 At 60 EeV $R_{50}$ is only 35\% of the proton energy loss length
 and would be less than a Mpc for Fe nuclei.  
  
 The magnetic field of our Galaxy is a good example for a combination
 of organized and random magnetic fields, which most likely exist on
 different scales in the Universe. The regular field $B_{reg}$ in the
 Galaxy has a spiral structure of axisymmetric or bisymmetric type
 resembling the matter distribution. The local strength of the field
 is about 1.8 $\mu$G with direction pointing inwards approximately
 along the Orion arm. The strength of the random field is not known
 exactly, with estimates between 1/2 and 2 $B_{reg}$. The correlation
 length $\ell$ of the random galactic fields is of order 50 to 100 pc.
 More  general estimates of the total field strength over the whole
 Galaxy give 5-6 $\mu$G~\cite{RBeck},
 and it is possible that a galactic halo field,
 that does  not contribute much locally, also exists. It is likely that
 the random field dominates the total field strength within the 
 galactic arms, while the regular field is dominant in the inter arm
 space.

 In a 5 $\mu$G galactic field the gyro radius of a 10$^{20}$ eV proton
 would be 20 kpc. The real deflection depends on the distance of
 the proton trajectory from the galactic center and on the pitch 
 angle of the trajectory to the regular field. 
 The Auger analysis attributes scattering angle of about 3$^o$ to
 their events above 60 EeV which often pass relatively close to the
 galactic center. This is consistent with the correlation angle 
 with nearby AGN measured by the collaboration. Events in the HiRes
 field of view will on the average smaller scattering than those 
 of Auger.

 The amount of scattering depends strongly on the exact magnetic field model.
 Field models with alternating polarity, such as BSS~\cite{Staapj}
 give the smallest deflections. Models with a $z$ component
 of the field (perpendicular to the galactic plane, that may be due
 to a dipole field) cause the largest ones.
 
  The possible existence of regular large scale fields makes the 
 consequences of proton propagation even more complicated.
 The following  exercise in Ref.~\cite{SSE} demonstrates the problems
 in the following geometry: a cosmic ray
 source at the origin injects isotropically protons above 10$^{18.5}$ eV
 on a power law spectrum with spectral index of 2 and exponential
 cutoff at 10$^{21.5}$ eV. The source is in the central $yz$ plane of
 a 3 Mpc wide magnetic wall, that is a simplified version of the
 Supergalactic plane (SGP), which is the plane of weight
 of galaxies within redshift of 0.04. Magnetic field with strength of
 $B_{reg}$ = 10 nG fills the SGP, points in $z$ direction and decays
 exponentially outside the SGP. The regular field is accompanied by
 random field with strength $B_{rndm} = B_{reg}/2$.

 Under these conditions the protons leaving the 20 Mpc sphere
 have not only very large deflections, but also modified energy
 spectra. Protons with trajectories perpendicular to the
 magnetic field lines exhibit much flatter spectrum since the lower 
 energy protons are caught up in the magnetic walls. Protons
 with trajectories parallel to the magnetic field lines have
 softer spectra that include the particles that could not penetrate
 the supergalactic plane. 
 
 In these simple cases one can scale the effects in proton energy
 as a function of the magnetic field strength. If $B_{reg}$ were 1 $\mu$G,
 for example, all effects would be the same in a 200 kpc sphere.
 There are lobes of radio galaxies, including Cen A, that have
 bigger lobes that may have ordered magnetic fields of that strength
 and may also cause significant deflections and modifications
 of the injection spectra depending on the position of the
 observer.

  \section{Production of secondary fluxes in propagation}

 The energy that UHECR lose in propagation ends up in fluxes of 
 gamma rays and neutrinos from $\pi^0$ and $\pi^\pm$ as well as
 other charged and neutral mesons decays.
 Gamma rays have even shorter energy loss length than nuclei
 and develop pair production/inverse Compton cascades where 
 synchrotron radiation may play important role. In the presence 
 of noticeable (1 nG) extragalactic field 10$^{18}$ eV electrons 
 quickly lose energy to GeV synchrotron photons.

 Neutrinos, however, have very low interaction cross section 
 and arrive at Earth with just redshift energy loss from
 almost any distance. For this reason the cosmological 
 evolution of the cosmic ray sources are very important input
 in the calculation of these {\em cosmogenic}~\cite{BerZat,Steck}
 neutrinos. 

\begin{figure}[thb]
\begin{center}
\includegraphics[width=7.5truecm]{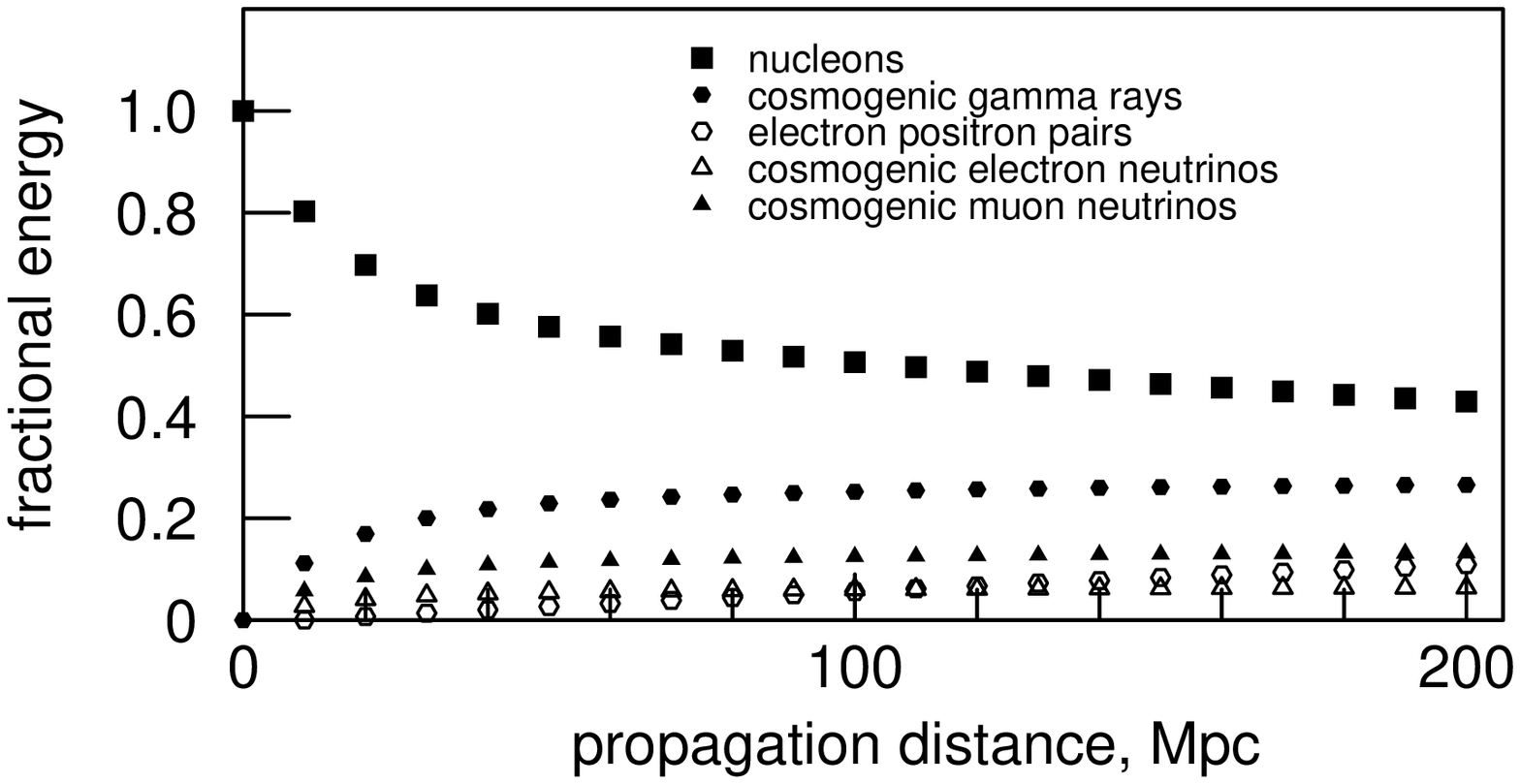}
\includegraphics[width=7.5truecm]{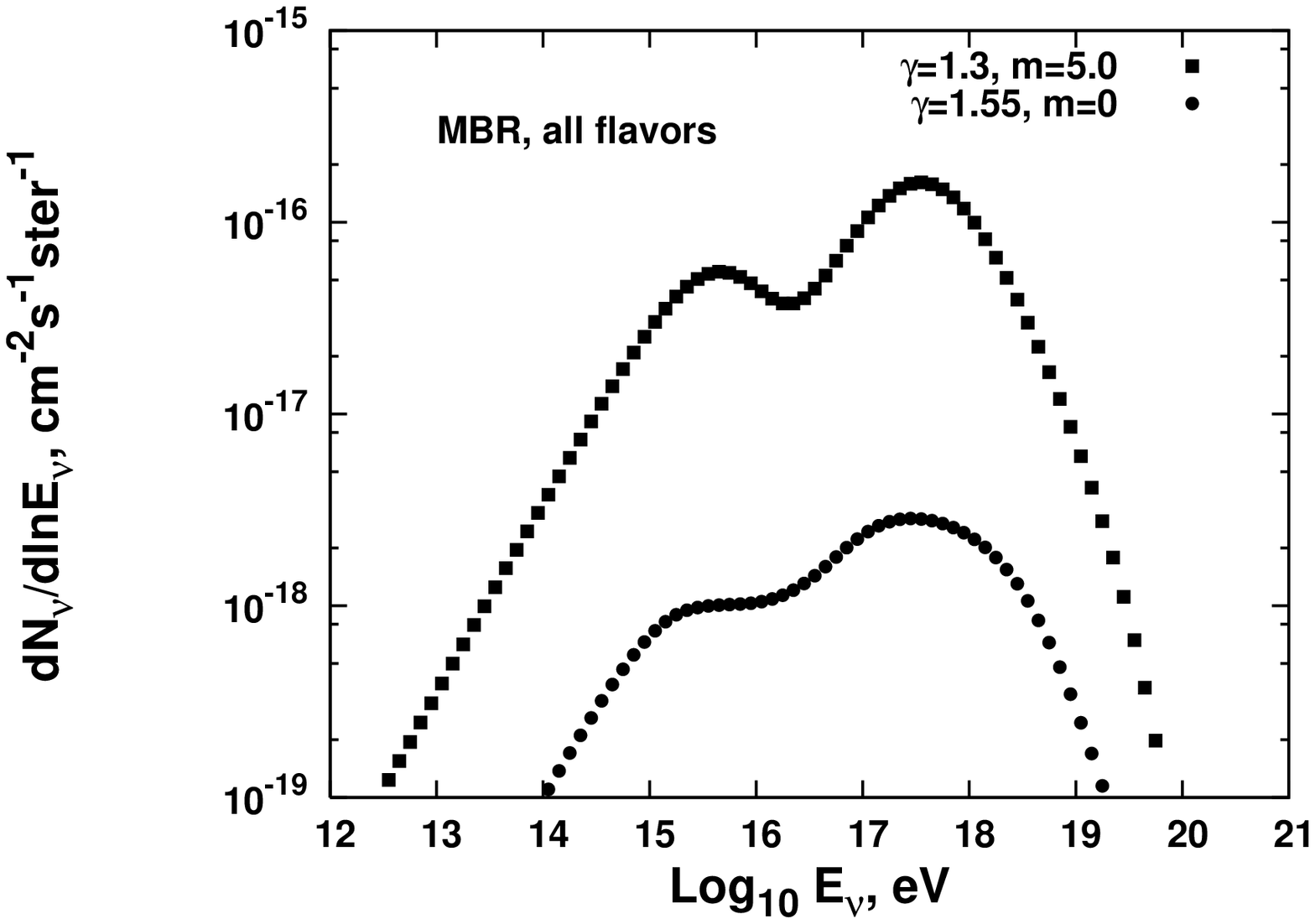}
\end{center}
\caption{
 Left hand panel: Fraction of the energy of the injected proton beam 
 above 10$^{19}$ eV ($\gamma$=1 spectrum) in nucleons and secondary 
 particles. 
  Right hand panel: Cosmogenic neutrinos from the two proton
 models that fit best the Auger cosmic rays spectrum. Note that 
 this is only the production in interactions on MBR. 
\label{cosm}
}
\end{figure}

 The left hand panel of Fig.~\ref{cosm} shows what fraction of the 
 injection energy of $E^{-2}$ cosmic ray spectrum above 10$^{19}$ 
 is contained in nucleons and secondary particles. After propagation
 on 200 Mpc less than 50\% of the primary energy content is in
 nucleons. The rest is distributed between neutrinos, gamma rays 
 and electron-positron pairs. The largest fraction of the energy loss
 is in gamma rays from neutral meson decays. The reason for this 
 is that the $\Delta^+$ decays to $p + \pi^0$ twice as often 
 as it decays to $n + \pi^+$. The muon neutrino flux consists 
 of approximately equal numbers of muon neutrinos and antineutrinos.
 The electron neutrino flux is mostly electron neutrinos. 

 The production of cosmogenic particle fluxes depends 
 on the same general astrophysical parameters as the 
 cosmic ray flux. The main difference is that the 
 cosmological evolution of the cosmic ray sources is 
 much more important. As we saw above UHE cosmic rays
 arrive only from very small redshifts, where even very strong
 cosmological evolution would make 10\% difference while the 
 cosmogenic neutrinos may easily come from $z$=2 which would
 correspond to an increase of order of magnitude or more.

 The infrared background (IRB) is also a target worth considering 
 when the production of cosmogenic neutrinos is discussed.
 Its number density is a factor of 400 or more less than 
 that of the MBR, but its energy is significantly higher.
 This means that lower energy nucleons will occasionally
 interact on IRB photons and generate lower energy neutrinos.
 Since even in a flat ($\gamma$=1) injection spectrum 
 there are many more lower energy protons the contribution
 of interactions in IRB to the cosmogenic 
 neutrino flux, although of lower energy, may be significant.
 The contribution of IRB  increases with the steepness of the
 UHECR spectrum.

 Two among the Auger results presented above may have an
 important connection to the flux of cosmogenic neutrinos -
 the correlation of the highest Auger events with nearby 
 AGN and the high cosmological evolution  
 required by one of the spectrum fits with relatively 
 flat  proton injection spectrum.

 The cosmological evolution of different types of astrophysical 
 objects is studied mostly in star forming regions (SFR) by
 infrared observations or in AGNs from their X-ray emission.
 The cosmological evolution of SFR is derived as $(1 + z)^3$
 on the average for up to redshifts close to 2. 

 It is indeed impossible to judge what is the best proxy for 
 the cosmic rays source evolution, infrared radiation or
 X-rays in the 0.5 - 2 KeV range where most of the statistics
 is. The conclusions from the X-ray observations are that 
 the evolution of AGN is close to $(1 + z)^5$~\cite{FSAD98,HMS05}.
 Very powerful AGN have even a stronger evolution but the large
 number of AGN correlating with the Auger UHECR require 
 using the evolution of average ones.
 
 It may be just a coincidence, but the Auger collaboration 
 analysis of the spectrum shown in Fig.~\ref{Aug_HR_prl}
 can be best described by two proton models~\cite{Yamamoto}: one with 
 $\gamma$ = 1.55 and no cosmological evolution, and 
 another on with $\gamma$=1.3 and cosmological evolution 
 $(1 + z)^5$, the same as that of AGN.
 The right hand panel of Fig.~\ref{cosm} shows the 
 cosmogenic neutrino spectra generated by these two models.
 The higher energy peak of these energy spectra consists 
 mostly of $\nu_\mu$, $\bar{\nu}_\mu$, and $\nu_e$ from 
 charged meson decays. The peak at about 10$^{15}$ eV consists
 only of $\bar{\nu}_e$ from neutron decay.
 The difference between the two neutrino models is huge, almost two
 orders of magnitude at maximum. One can also see the redshift
 of the maximum in the $\gamma$=1.3 model where most of the neutrinos
 are generated at high redshift. This also shows up in the lower
 energy $\bar{\nu}_e$ peak that is enhanced because of the 
 higher number of protons (and respectively secondary neutrons)
 injected at high redshifts. The $\gamma$=1.55 model flux 
 will significantly increase if the production in IRB is included.
 This would, however, be at lower energy and the detection rate will
 not change a lot.

 The two other models that fit the Auger spectrum are mixed composition
 ones. If UHECR are heavy nuclei they generate a few cosmogenic neutrinos
 and only the proton fraction is efficient in production of meson 
 decay neutrinos. Since there are many neutrons that are released in
 the spallation process there is a significant flux of $\bar{\nu}_e$
 from neutron decay. There are several
 calculations~\cite{heavnu}
 of the cosmogenic neutrino fluxes from mixed composition models that 
 show this effect.

 A possible detection even of a few cosmogenic neutrinos would 
 greatly help the interpretation of the detected UHECR fluxes. 
 Because of the big difference in the neutrino flux only models 
 with strong cosmological evolution may lead to detection. 
 For those without cosmological evolution we need Gigaton
 neutrino detectors.

\section{Summary}

\noindent
\hspace*{36pt}$\bullet$ All known stable particles that are candidates for the
 ultra high energy cosmic rays lose energy in interactions with
 the cosmic microwave background and other astrophysical photon fields.

\noindent
\hspace*{36pt} The cosmic rays spectrum measured by the HiRes collaboration
 agrees very well with a model of pure proton composition while the 
 Auger collaboration spectrum may require mixed primary composition.

\noindent
\hspace*{36pt}$\bullet$ The energy loss sets a horizon for the sources of such
 particles (a maximum distance from the observer) and modify the 
 injection spectrum of these cosmic rays.

\noindent
\hspace*{36pt}$\bullet$ Charged UHECR, such as protons and heavier nuclei,
 scatter
 in the extragalactic magnetic fields. This scattering causes increased
 pathlength, decreases the particle horizon, and introduces deflection
 from the source direction and significant time delay.

\noindent
\hspace*{36pt}$\bullet$ The cosmogenic neutrino fluxes generated in cosmic rays
 propagation are currently close to detection in the km$^3$ detectors
 if the models with strong cosmological evolution of UHECR are correct.

\noindent
\hspace*{36pt}$\bullet$ Only protons of energy approaching 10$^{20}$ eV reveal
 the source position and spectrum after an account for the energy loss
 and scattering in propagation.

\noindent
\hspace*{36pt}$\bullet$ The identification of the UHECR sources will bring
 very valuable general information about the power and size of the
 sources and the magnetic fields in them, as well in the intergalactic 
 space.  
 \vspace*{3truemm}

\noindent
{\bf Acknowledgments} Much of the work on which this paper is based
 was performed in collaboration with R.~Engel, D.~Seckel and others.

\end{document}